\documentclass[10pt]{iopart}
\usepackage{graphics}
\usepackage{iopams}

\usepackage{subfigure}
\usepackage{color}
\begin{document}

\title[Non-gapless excitation and zero-bias fast oscillations ...]
{Non-gapless excitation and zero-bias fast oscillations in the LDOS of surface superconducting states}

\author{Liangyuan Chen$^{1,2}$, Yajiang Chen$^{1,*}$, ZHANG Wenhui$^{\dagger, 3}$, ZHOU Shuhua$^{4}$}
\address{1. Key Laboratory of Optical Field Manipulation of Zhejiang Province, Department of Physics, Zhejiang Sci-Tech University, 310018 Zhejiang, China}
\address{2. College of Engineering, Lishui University, 323000, Zhejiang, P. R. China}
\address{3. School of Electronic Engineering, Nanjing Xiaozhuang University, Nanjing, China}
\address{4. Zhejiang Economic Vocational and Technical College, Hangzhou, China}
\ead{* yjchen@lsu.edu.cn; $\dagger$ hit\_zwh@126.com}
\date{\today}

\begin{abstract}
Recently a novel surface pair-density-wave (PDW) superconducting state has been discovered in Refs. [Phys. Rev. Lett. \textbf{122}, 165302 (2019)] and Phys. Rev. B \textbf{101}, 054506 (2020)], which may go through a distinct multiple phase transition (MPT) when the superconductivity fades away from bulk to the boundary (e.g. edges and corners). Based on the Bogoliubov-de Gennes equations for the attractive tight-binding Hubbard modal in a one-dimensional chain, we demonstrate that the surface PDW state has a non-gapless quasiparticle spectrum, which is contrary to the conventional surface superconducting state. Moreover, we find that the MPT is associated with a zero-bias fast oscillating pattern in the LDOS near the surface. Our findings provide a potential experimental clue to identify the surface PDW state.
\end{abstract}

\maketitle
\section{Introduction}\label{int}
Surface superconductivity has been intensively studied for decades due to its theoretical and experimental importance for finite superconductors and identification of superconducting mechanism. It is known that~\cite{SaintJames1963PL} surface superconductivity can survive at the metal-insulator interface of a infinite bulk cylinder with thickness about $\xi(T)$ under a parallel magnetic field $H$ between the bulk upper critical field $H_{c2}$ and the nucleation field $H_{c3}$ ($H_{c3} =1.69 H_{c2})$, which has been confirmed by experiments for both type-$I$ and $II$ superconductors~\cite{Hempstead1964PRL,Strongin1964PRL}. For mesoscopic samples $H_{c3}$ increases~\cite{Schweigert1998PRB,Brosens1999SSC} due to the enhanced quantum confinement~\cite{Chen2010,Chen2012a}. Various interesting properties are relevant to surface superconductivity. For example, iron-based superconductor FeTe$_{1-x}$Se$_x$(x=0.45) hosts Dirac-cone-type spin-helical superconducting surface states with an $s$-wave gap~\cite{Zhang2018S}. The electric-field-induced surface superconductivity in SrTiO$_3$ shows a multiple-gap structure\cite{Mizohata2013PRB}. It even broadens the superconducting transition of Nb films~\cite{Zeinali2016PRB}. Surface superconductivity has been discovered in many other materials, e.g. MgB$_2$~\cite{Rydh2003PRB,Welp2003PRB}, noncentrosymmetric PbTaSe$_2$~\cite{Guan2016SA}, metal-Dirac-semimetal interface in Cd$_3$As$_2$~\cite{Huang2019NC}, Weyl loop materials~\cite{Wang2017PRB} and Fulde-Ferrell-Larkin-Ovchinnikov (FFLO) superconductors~\cite{Samokhin2019PRB}.

Recently, novel surface pair-density-wave (PDW) superconducting and superfluid states have been predicted theoretically~\cite{Barkman2019PRL} in materials supporting the FFLO state at a sufficient large Zeeman splitting field, which plays a role of making a fermionic population imbalance~\cite{Radzihovsky2011PRA,Zwierlein2006S}. The existence of surface PDW state relies on the negative energy density at the boundary according to the analysis based on the extended Ginzburg-Landau (GL) theory, along with an exponential enhancement of the order parameter close to the boundary~\cite{Samoilenka2020PRB}. According to the microscopic Bogoliubov-de Gennes (BdG) calculations~\cite{croitoru2020}, the electron kinetic energy in the single-particle Hamiltonian plays the primary role for the surface PDW state, rather than the Cooper-pair interactions or the lattice potential. Thus, the surface PDW state here differs from the PDW state, which roots in the non-conventional interaction between paired electrons in various superconducting materials, e.g. cuprates~\cite{Du2020}, Kagome metals~\cite{Chen2021}, transition-metal dichalcogenides~\cite{Liu2021}.

Moreover, the surface PDW state can exist even without the magnetic field, resulted by the constructive interference~\cite{croitoru2020} between the quasiparticles near the surface. The numerical calculations in Ref.~\cite{Barkman2019PRL} show that for one-dimensional systems the surface PDW state is more robust than the FFLO state. An interesting property of the surface PDW state is the multiple phase transition (MPT)~\cite{Samoilenka2020PRB} from the superconducting state to the normal state, which occurs separately in space from bulk to the boundary (e.g. edges and corners). Such MPT behavior has been predicted in the frame of the generalized GL theory~\cite{Buzdin1997PLA} with higher order derivative terms, and the microscopic Bogoliubov-de Gennes (BdG) equations with the attractive tight-binding Hubbard model~\cite{Radzihovsky2011PRA,Micnas1990RMP,Huscroft1997PRB}.

In the microscopic view, the properties of both the conventional surface superconducting state and the surface PDW state are directly determined by the behavior of Bogoliubon quasiparticles~\cite{Gennes1966}. From literature, we learn that the quasiparticle energy spectrum of the conventional surface superconducting state is gapless under $H_{c2} < H < H_{c3}$, because the quasiparticles of the normal (inner) region can infiltrate into the surface superconducting region, which has been confirmed by the tunneling experiments. It is interesting to study whether the quasiparticle energy spectrum of the surface PDW state is also gapless. However, this issue still remains unclear. 

In experiments, the information of the quansiparticles in the surface superconducting state can be explored commonly by the local density of states (LDOS) obtained by scanning tunneling microscopy (STM). It has been found that the surface LDOS spectrum of $s$-wave surface superconducting states shows a significant depletion near the Fermi energy caused by a local superconducting order~\cite{Troy1995PRB}. The topological surface superconducting states in PbTaSe$_2$ strongly modulate~\cite{Guan2016SA} the LDOS mapping in real space. In the presence of the Majorana edge states (i.e. Andreev bound state) in $d_{xy}+p$-wave superconductors, the LDOS spectrum at the surface significantly depends on the direction of the Zeeman field~\cite{Yada2011PRB}. For $p_x \pm i p_y$-wave superconductors, quasiparticle states lead to a flat LDOS at the surface~\cite{Matsumoto1999JotPSoJ} with no peak-like or gap-like structure within the energy gap. The LDOS of FFLO states in $d$-wave superconductors exhibits~\cite{Vorontsov2005PRB} Andreev resonance which can be shifted by an applied magnetic field. All quasiparticle resonances locate in the region where the order parameter is suppressed. In addition, the LDOS spectrum has been regarded as a microscopic evidence to distinguish one-dimensional and two-dimensional FFLO states in $d$-wave superconductors~\cite{Wang2006PRL,Zhou2009PRB}. 

The feature in the LDOS of the surface PDW superconducting state can be a solid experimental clue to verify the presence of such state. The pioneer study on the LDOS spectrum was given for a one-dimensional chain~\cite{Samoilenka2020PRBa} in the bulk-temperature ($T$) regime. At such low $T$, the superconducting order parameter is nonzero around the sample center, thus, unfortunately, the surface PDW state is strongly masked. However, it is reasonable to expected that surface PDW state possesses distinct features in the surface LDOS spectrum as compared with that in the bulk, especially for the MPT in a square system~\cite{Samoilenka2020PRB}. In the present work we investigate the minimal quasiparticle energy and the LDOS of the surface PDW state in a one-dimensional chain by numerically solving the BdG equations with the attractive tight-binding Hubbard model in a wide range of $T$ and Debye window $\hbar\omega_D$. Surprisingly, unlike the conventional surface superconducting state, the minimal quasiparticle energy spectrum $E_{\rm min}$ of a surface PDW state is non-zero, i.e. non-gapless, even when the order parameter in the center vanishes.

The paper is organized as follows. In Sec.~\ref{sec2} we outline the formalism of the BdG equations with the tight-binding Hubbard model in a one-dimensional chain. Sec.~\ref{sec3} presents the numerical results of the minimal quasiparticle energy and the LDOS spectra for a one-dimensional chain. Our results are summarized in Sec.~\ref{sec4}.

\section{Theoretical formalism}\label{sec2}
The Bogoliubov-de Gennes equations for a tight-binding model with an on-site attractive coupling in a one-dimensional chain can be written as follows~\cite{Samoilenka2020PRB,Tanaka2000PRB}
\begin{eqnarray}\label{bdg}
E_\nu u_\nu(i) &= \Sigma_{i'} A_{ii'} u_\nu(i') + \Delta_i v_\nu(i) \\
E_\nu v_\nu(i) &= \Delta_i^* u_\nu(i) -\Sigma_{i'} A_{ii'} v_\nu(i'),
\end{eqnarray}
where $\{E_\nu,\, u_\nu(i),\, v_\nu(i)\}$ are the quasiparticle energy, electron- and hole-like wavefunctions, respectively. The Hartree-Fock potential has been ignored in our calculation since it gives a confinement potential which only produces a shift of the quantum-size oscillations.~\cite{chen2009} And, the matrix $\{A_{ii'}\}$ in the absence of the magnetic field is of the form
\begin{equation}\label{a}
A_{ii'}=-\Sigma_\delta t_{\delta}(\delta_{i',i-\delta} + \delta_{i',i+\delta})-\mu\delta_{ii'}
\end{equation}
with $t_\delta$ the hopping rate. For simplicity, we restrict ourselves to the nearest-neighbour hopping, i.e. $\delta=1$. The number of sites in the chain is taken as $N=181$, which has been verified to be sufficiently large to overcome the quantum-size effect. The chemical potential $\mu$  is determined by the electron filling level $n_e$ at each site:
\begin{equation}\label{ne}
n_e = \frac{2}{N}\Sigma_{\nu,i}\bigg[ f_\nu |u_\nu(i)|^2 + (1-f_\nu) |v_\nu(i)|^2\bigg], 
\end{equation}
where $f_\nu=f(E_{\nu})$ is the Fermi-Dirac distribution. In our calculation we focus on the half-filling situation $n_e=1$, which steadily guarantees the presence of the surface PDW state~\cite{Samoilenka2020PRBa}. The expression of the order parameter $\Delta_i$ is the self-consistent condition, and can be rewritten as 
\begin{equation}\label{op}
\Delta_i = g\Sigma_\nu \big[ u_\nu(i)v^*_\nu(i)(1-2f_\nu) \big],
\end{equation} 
where $g$ is the coupling strength, and the summation should be limited to the quasiparticle states with $0<E_\nu<\hbar\omega_D$ ($\hbar\omega_D$ is the Debye window around $\mu$).

The BdG Eqs.~(\ref{bdg}) are solved self-consistently together with Eq.~(\ref{op}). Firstly, construct a discrete numerical BdG matrix with an initial value for $\Delta_i$ according to Eq.~(\ref{bdg}). Secondly, calculate the quasiparticle wavefunctions and energy spectrum by diagonalizing the numerical BdG matrix. Thirdly, obtain new $\Delta_i$ by inserting quasiparticle information into Eq.~(\ref{op}). Repeat these steps until the difference between the new and old order parameters is sufficiently small, i.e. $\Delta_i$ converges. Moreover, the local density of states (LDOS) is given by 
\begin{eqnarray}\label{ldos}
	\rho_i(V) \propto -\Sigma_{\nu} \big[ |u_\nu(i)|^2f'(E_\nu-eV) +  |v_\nu(i)|^2f'(E_\nu+eV) \big],
\end{eqnarray}
while the total density of states (DOS) is 
\begin{equation}\label{dos}
 	D(V) \propto \Sigma_i \rho_i(V).
 \end{equation}
 
It is of importance to notice that the BdG equations are valid in the whole regime of $T$ and the external magnetic field, while near the transition point it goes into the form of Ginzburg-Landau theory, which normally works in the vicinity of the transition point. In the numerical calculations, we set the lattice constant of the one-dimension chain to be unit, and take the energy quantities (e.g. $E_\nu$, $\Delta_i$, $\hbar\omega_D$, $\mu$) in the unit of the hopping rate $t$. Therefore, $g$ in Eq.~(\ref{op}) is also in unit of $t$.

\section{Results and discussions}\label{sec3}
\begin{figure}[!h]
\center
\resizebox{0.8\columnwidth}{!}{\rotatebox{0}{\includegraphics{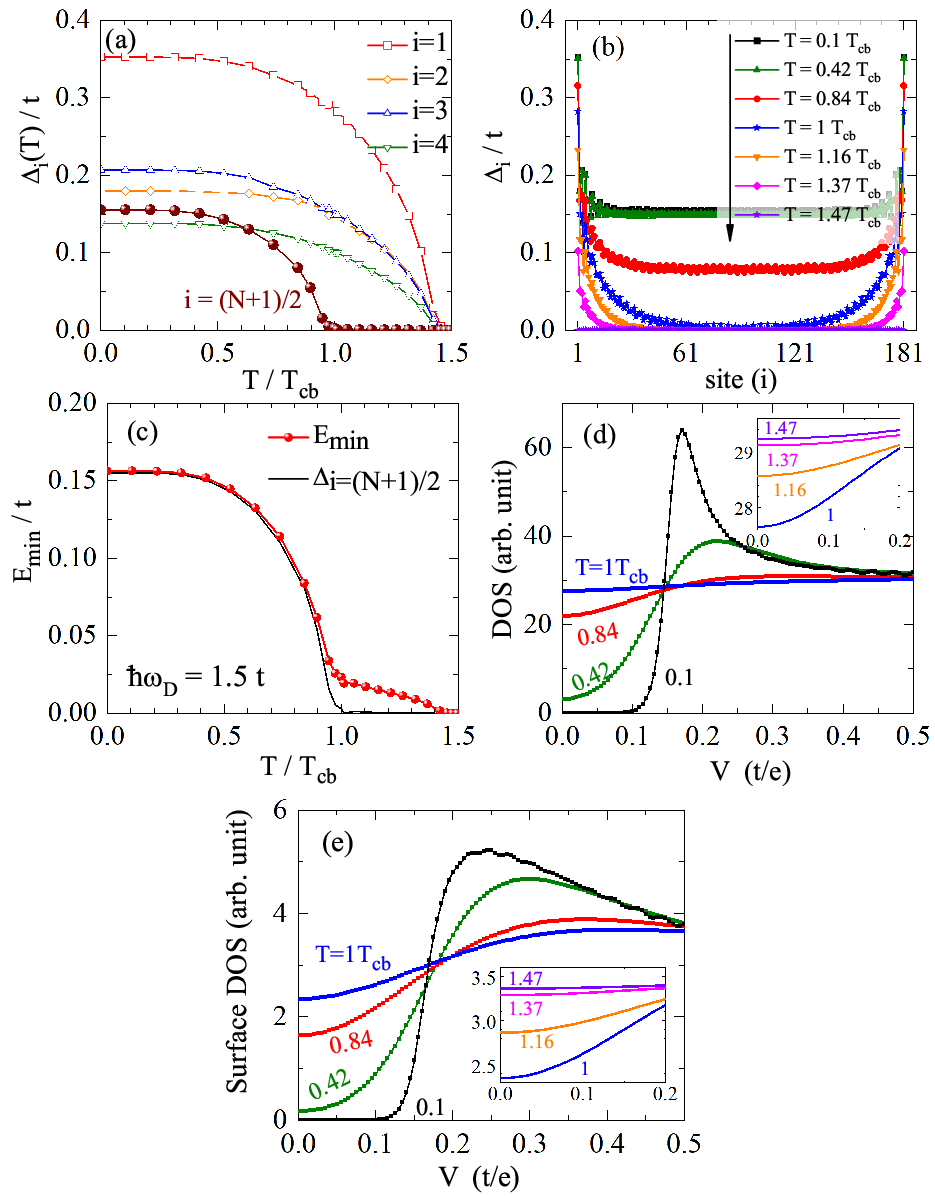}}}
\caption{(Color online) (a) The order parameter $\Delta_i$ as a function of $T$ in the unit of the bulk critical temperature $T_{cb}$ for $\hbar\omega_D=1.5\,t$, $g=2\,t$ and $N=181$. The sites $i=1,\,2,\,3,\,4$ and $(N+1)/2$. (b) Spatial order parameter $\Delta_i$ at different $T$ values. (c) Minimal quasiparticle energy $E_{min}$ as a function of $T$. The solid curve shows $\Delta_{i=(N+1)/2}(T)$ as a reference. (d) The DOS spectrum versus the bias $V$ at various $T$, while the insert panel shows the results at $T>T_{cb}$. (e) The surface DOS corresponds to the same temperatures that only includes the surface sites $i \le N_s $ and $i \ge N-N_s$ with $N_e = 10$, at which $\Delta_i$ decays to about $10\%\,\Delta_{i=1}$.}  \label{fig1}
\end{figure}

\begin{figure}[!h]
\center
\resizebox{0.8\columnwidth}{!}{\rotatebox{0}{\includegraphics{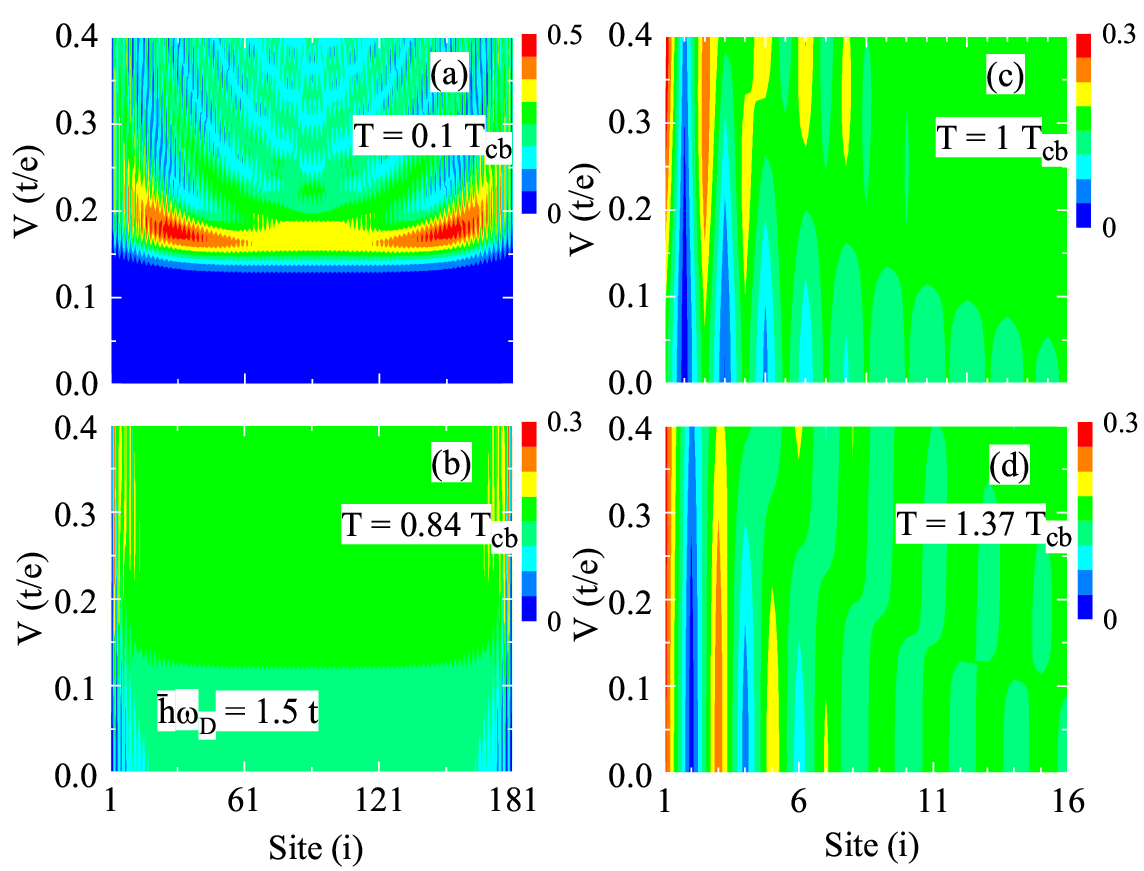}}}
\caption{(Color online) The LDOS contour plots for $\hbar\omega_D=1.5 t$, $g=2\,t$ and $N=181$ at $T=0.1\, T_{cb}$ (a), $T=0.84\, T_{cb}$ (b), $T=1\, T_{cb}$ (c) and $T=1.37\, T_{cb}$ (d).} \label{fig2}
\end{figure}

\begin{figure}[!h]
\center
\resizebox{0.8\columnwidth}{!}{\rotatebox{0}{\includegraphics{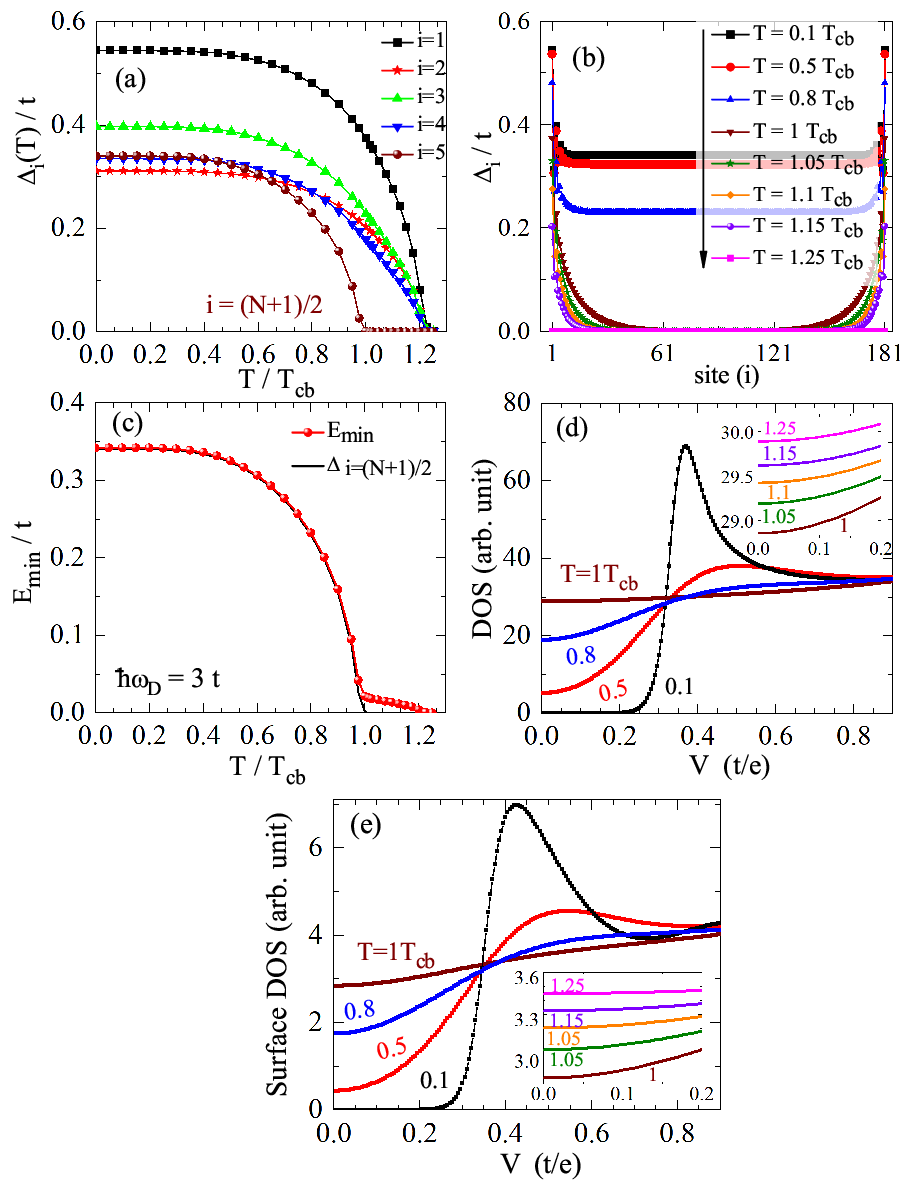}}}
\caption{(Color online) The same with Fig.~\ref{fig1}, but for $\hbar\omega_D=3\,t$, $g=2\,t$, $N=181$ and $N_s=15$.} \label{fig3}
\end{figure}

\begin{figure}[!h]
\center
\resizebox{0.8\columnwidth}{!}{\rotatebox{0}{\includegraphics{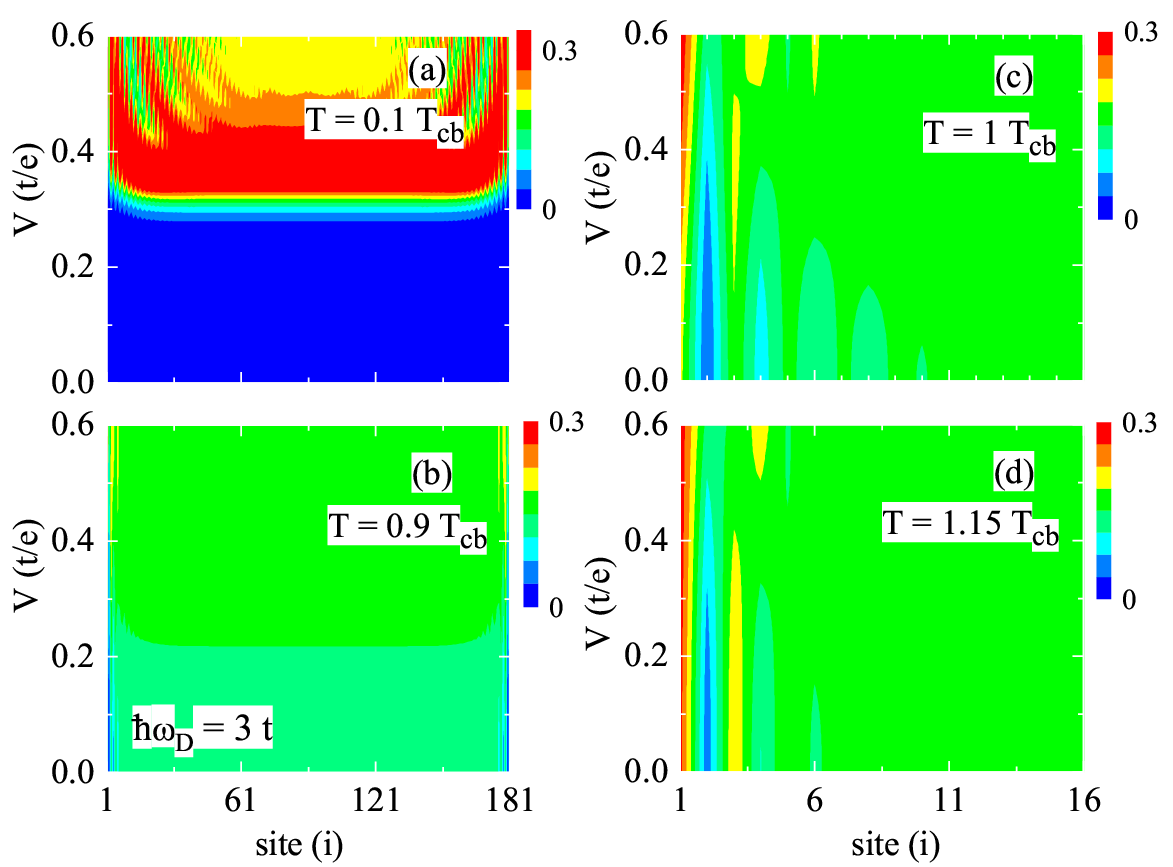}}}
\caption{(Color online) The same with Fig.~\ref{fig2}, but for $\hbar\omega_D=3\,t$, $g=2\,t$ and $N=181$.} \label{fig4}
\end{figure}

Fig.~\ref{fig1}(a) shows the spatial order parameter $\Delta_i$ as a function of $T$ close to the surface ($i=1$, $2$, $3$, $4$), and at the center of the chain $i=(N+1)/2$. The following parameters are applied in the calculations: $\hbar\omega_D=1.5\,t$, $g=2\,t$ and $N=181$. The $T$-dependence of $\Delta_i$ for the whole chain is given in Fig.~\ref{fig1}(b). It is clearly seen that the superconductivity of this sample is destroyed at the center of the chain as $T$ increases up to $1.0\, T_{cb}$, while the surface PDW superconductivity survives at $T < T_{cs}=1.45\,T_{cb}$. For the given parameters, $T_{cb}=0.1\, t/k_B$ with $k_B$ the Boltzman constant. This agrees with the results in Refs.~\cite{croitoru2020,Samoilenka2020PRBa}. The critical temperature of the surface PDW state $T_{cs}$ is about $45\%$ higher than the bulk critical temperature $T_{cb}$. The minimal quasiparticle energy $E_{\rm min}$ (the red curve with symbols) is given in Fig.~\ref{fig1}(c) as a function of $T$, while the black curve represents $\Delta_{i=(N+1)/2}$ as a reference. It is known that for the conventional surface superconducting state, the quasiparticle energy spectrum is gapless as introduced in Sec.~\ref{int}. However, surprisingly, one can find that $E_{\rm min}=0.023\,t$ at $T=T_{cb}$, which is $14.8\%$ of the zero-$T$ $E_{min}$. In addition, $E_{min}$ is nonzero at $T_{cb}<T< T_{cs}$, and vanishes at $T=T_{cs}$. It means that the surface PDW superconducting state has a non-gapless quasiparticle energy spectrum, which is contrary to the case of the conventional surface superconducting state. The DOS calculated by Eq.~(\ref{dos}) is plotted as a function of the bias $V$ in Fig.~\ref{fig1}(d). The unit of $V$ is $t/e$ with $e$ the elementary charge. The non-gapless quasiparticle energy spectrum causes a slight drop of the DOS close to $V=0$ at $T_{cb} < T < T_{cs}$. 
One may notice that the bias value, that corresponds to the maximum of the DOS curve, grows as increasing $T$. This is due to the thermal pair-breaking effect according to the Dynes' model~\cite{chen2019}, the similar behavior has been observed experimentally in superconducting nanoscale islands~\cite{Brihuega2011}.

In Fig.~\ref{fig1}(e) we show the DOS for the surface region $i<N_s$ and $i>N-N_s$, i.e. integration of the local density over the surface sites. Here, $N_s$ is taken as $N_s=10$, at which $\Delta_i$ decays to about $10\%\,\Delta_{i=1}$. One can find that the surface DOS is much less than the total value at all regions of the bias and $T$. But at $T_{cb}< T < T_{cs}$ the increasement of the total DOS at zero bias mostly comes from that of the surface DOS, which reveals that the surface plays a primary role when $T$ increases and goes close to $T_{cs}$.

The contour plots of the LDOS for this sample obtained by Eq.~(\ref{ldos}) are illustrated in Fig.~\ref{fig2} (a-d) at $T/T_{cb}=0.1$, $0.84$, $1$ and $1.37$, respectively. In all of these plots the ranges of $V$ are set as $[0,\,0.4\,t/e]$, while the color bars in plots (b-d) are the same. At $T=0.1\,T_{cb}$, from Fig.~\ref{fig2}(a) we find that along the chain there is a global gap at $V<0.145\, t/e$ along with a fast oscillation pattern, which matches the results in Ref.~\cite{Samoilenka2020PRB}. Apparently, these two features are relevant to the whole spatial variation of $\Delta_i$ as seen in Fig.~\ref{fig1}(b). At such low $T$, the surface PDW state is strongly concealed by the bulk behavior of $\Delta_i$. When $T=0.84\,T_{cb}$ as in Fig.~\ref{fig2}(b), the global gap in the LDOS spectrum has been smeared out by thermal effect, and the spatial oscillating pattern shifts to zero bias since the surface PDW state becomes more significant. In Fig.~\ref{fig2}(c), the oscillation pattern splits into two branches at $T=1\,T_{cb}$: one decreases toward to $V=0$, the other toward to higher $V$. The periods of both oscillations at zero bias are about $2$ sites. At $T=1.37\,T_{cb}$, the oscillation amplitude of the first branch is enhanced. Such oscillation pattern is controlled by the quasiparticles with low energy, and can be employed as the experimental signal of the surface PDW state.

Fig.~\ref{fig3} gives $\Delta_i(T)$, $E_{\rm min}$ and DOS of the surface PDW state for $g=2\,t$, $N=181$ and $\hbar\omega_D=3\,t$. As seen in Fig.~\ref{fig3}(a) and (b), the system shows the surface PDW state. In this case, $T_{cb}=0.20\,t/k_B$ and $T_{cs}=0.245\,t/k_B$. In Fig.~\ref{fig3}(c), one can find that $E_{\rm min}$ at $T=T_{cb}$ is equal to $0.0214\,t$. It is $6.3\%$ of $E_{\rm min}$ at $T=0\, T_{cb}$, which is less than the ratio $14.8\%$ in Fig.~\ref{fig1}(c). In the regime of $T_{cb} < T < T_{cs}$, this percentage is depressed when compared with the case in Fig.~\ref{fig1}(c). As a result of this non-gapless quasiparticle energy spectrum, the DOS at $T_{cb} < T < T_{cs}$ in the inner plot of Fig.~\ref{fig3}(d) exhibits a minimum at $V=0$, which is similar to the case in Fig.~\ref{fig1}(d).  
As the same in Fig.~\ref{fig1}(e), the surface DOS spectrum is illustrated in panel (e). Although the surface quantity are greatly less than the total value, their profiles are similar. Moreover, when comparing with panels (d) and (e), we can find that the increase of the total DOS at zero bias mainly comes from the surface DOS, which can also be a signature of the existence of the surface PDW state.

The LDOS plots are shown in Fig.~\ref{fig4}(a-d) for $T/T_{cb}=0.1$, $0.9$, $1$ and $1.15$, respectively. All the color maps are unified in these plots, while $0\le V \le0.6 \,t$. Similar to Fig.~\ref{fig2}, we find a global gap along the spatial direction, which is about $0.3\,t$. This gap is enhanced at about $10$ sites near the surface, and is smeared out at $T=0.9\,T_{cb}$ in Fig.~\ref{fig4}(b). When $T$ increases up to $T_{cb}$, a fast oscillations show up at zero bias near the surface ($i<5$). However, the oscillating strength is much weaker than that in Fig.~\ref{fig2}(c). For $T=1.15\,T_{cb}$, at which only $\Delta_{i}$ close to the surface is non-zero, the fast oscillation pattern can be still clearly identified. 

\section{Conclusions}\label{sec4}
By numerically solving the Bogoliubov-de Gennes equations for the attractive tight-binding Hubbard modal in a one-dimensional chain, the recently discovered surface pair-density-wave (PDW) state has a non-gapless quasiparticle spectrum, which is opposite to the case of the conventional surface superconducting state sustained under a magnetic field between $H_{c2}$ and $H_{c3}$. The multiple phase transition from the bulk superconducting state to the surface PDW state is associated with a fast oscillation pattern in the LDOS spectrum at zero bias near the surface, which can be applied as a potential experimental clue to identify the surface PDW state.

\section*{ACKNOWLEDGMENTS}
This work was supported by Open Foundation of Key Laboratory of Optical Field Manipulation of Zhejiang Province(ZJOFM-2020-007), Science Foundation of Zhejiang Sci-Tech University(ZSTU) (Grant No. 19062463-Y),  Natural Science Foundation of Zhejiang Province (Grants No. LY18A040002, LY18F030001 and LY20E050002), Key projects of Natural Science Foundation of Zhejiang Province(LZ21F020003) and Nanjing Xiaozhuang College Talent Fund (2020 NXY14).


\end{document}